\documentclass[aip,amsmath,amssymb,reprint,onecolumn]{revtex4-1}

\usepackage{graphicx}
\usepackage{dcolumn}
\usepackage{bm}
\usepackage[utf8]{inputenc}
\usepackage[T1]{fontenc}
\usepackage{mathptmx}
\usepackage{etoolbox}

\makeatletter
\def\@email#1#2{%
 \endgroup
 \patchcmd{\titleblock@produce}
  {\frontmatter@RRAPformat}
  {\frontmatter@RRAPformat{\produce@RRAP{*#1\href{mailto:#2}{#2}}}\frontmatter@RRAPformat}
  {}{}
}%
\makeatother
\begin{document}
\preprint{AIP/123-QED}

\title{Time-reversal positivity}

\author{Shen-Hao Xu}
\affiliation{Tianjin University of Technology, Tianjin 300384, China}
%orcid{0009-0009-7186-9321}

\author{Zhong-Chao Wei}
\email{zcwei@thp.uni-koeln.de}
\affiliation{Qingdao Binhai Univeristy, Qingdao 266555, China}
%orcid{0000-0002-1708-4306}

\date{\today}
\begin{abstract}
We propose a new analytical tool called time-reversal positivity.
It is an analogue of the Majorana reflection positivity in time-reversal
symmetric case. This new time-reversal positivity can fully explain
the relationship between time-reversal symmetry and the sign-free
property in quantum Monte Carlo simulations. As an application,
using a cone-theoretical method, we show the ground state uniqueness
for the time-reversal symmetric Hubbard model.
\end{abstract}
\maketitle

\section{Introduction}

Analytical tools to study interacting quantum lattice models are of
great value in fundamental research of physics. The uniqueness of
the ground states is a common research topic. This property has been
proved for the Heisenberg model\citep{lieb_ordering_1962} based on
the Perron-Frobenius theorem for matrices with non-negative elements.
For the fermionic Hubbard model, Elliott H. Lieb proposed a famous
method based on reflection positivity\citep{lieb_two_1989}. The reflection
positivity is actually a property about completely positive maps and
is an important property of operators and quantum states\citep{kondratiev_reflection_2006,biskup_reflection_2009}.
It was then generalized to the Majorana reflection positivity\citep{jaffe_reflection_2015,jaffe_characterization_2016},
and the ground state uniquness for an interacting spinless fermion
model was proved based on it\citep{wei_ground_2015}. Recently, using
the Perron-Frobenius theorem for convex cones\citep{tam_matrices_2006},
the uniqueness of the ground states of two Bose Hubbard models has
been proved\citep{wei_liebs_2025}. The proof is based on the Fock
space positivity and the Fock space reflection positivity.

In this work, we introduce two new interrelated research tools called
the time-reversal positivity, which is a property of time-reversal
symmetric operators, and the Fock space time-reversal positivity,
which is a property of time-reversal symmetric states. They can be
seen as analogues of the Majorana reflection positivity, and the major difference
is that they are about time-reversal symmetry instead of reflection
symmetry.

Positivities like the Majorana reflection positivity is known to have
a close relationship with the sign-free property of quantum Monte
Carlo simulations\citep{wei_majorana_2016}. So do the Fock space
positivity and the Fock space reflection positivity\citep{wei_liebs_2025}.
The relationship between time-reversal symmetry and the sign-free
property has long been a mystery. The new positivities in this work
can fill this gap.

Using the Fock space time-reversal positivity, we generalize Lieb's
theorem on the uniqueness of the ground states for the fermionic Hubbard
model to time-reversal symmetric case.

\section{Time-reversal positivity and Fock space time-reversal positivity}

Let $c_{i}$, $c_{i}^{+}$and $d_{i}$, $d_{i}^{+}$($i=1,\dots,L$)
be two sets of fermion annihilation and creation operators, which
satisfy the canonical anti-commutation relations: $\left\{ c_{r},c_{s}^{+}\right\} =\left\{ d_{r},d_{s}^{+}\right\} =\delta_{rs}$,
$\left\{ c_{r},d_{s}^{+}\right\} =\left\{ d_{r},c_{s}^{+}\right\} =0$,
$\left\{ c_{r},c_{s}\right\} =\left\{ c_{r}^{+},c_{s}^{+}\right\} =\left\{ d_{r},d_{s}\right\} =\left\{ d_{r}^{+},d_{s}^{+}\right\} =0$.
Denote $f^{+}=\left(c_{1}^{+},\dots c_{L}^{+},d_{1}^{+},\dots d_{L}^{+}\right)$,
and for any $k\in\mathbb{C}^{2L}$ denote $f_{k}^{+}=f^{+}\cdot k$.
For any $k_{1},\dots,k_{s}\in\mathbb{C}^{2L}$ denote the quantum
state $f_{k_{1}}^{+}\dots f_{k_{s}}^{+}\left|0\right\rangle $ as
$\left|k_{1}\dots k_{s}\right\rangle $.

Define the time-reversal operation $T$ by $T\left(k\right)=\left(-i\sigma_{y}\otimes I_{L}\right)\bar{k}$
and $T\left(f_{k}^{+}\right)=f_{T\left(k\right)}^{+}$, $T\left(f_{k}\right)=f_{T\left(k\right)}$,
where $\sigma_{y}$ is the Pauli matrix. It is is an anti-linear transformation
that sends $c_{s}\mapsto d_{s}$, $d_{s}\mapsto-c_{s}$ and $i\mapsto-i$.

In the linear space of fermion parity odd operators, $T^{2}=-1$.
In the linear space of fermion parity even operators, $T^{2}=1$.
Consider the real vector space $V_{O}$ spanned by time-reversal symmetric
operators $\hat{O}_{s}$ with even fermion parity, $T\left(\hat{O}_{s}\right)=\hat{O}_{s}$.
The whole complex linear space of fermion parity even operators can
be seen as a complexification of $V_{O}$. $V_{O}$ is spanned by
all the operators of the form $\chi_{1}T\left(\chi_{1}\right)\dots\chi_{s}T\left(\chi_{s}\right)\dots\chi_{k}T\left(\chi_{k}\right)$,
where $\chi_{s}=f_{k_{s}}$ or $f_{k_{s}}^{+}$.

Consider the convex cone $K_{O}\subseteq V_{O}$ generated(including
the closure) by the time-reversal invariant operators of the following
type:
\begin{equation}
\left(-1\right)^{r}\chi_{1}T\left(\chi_{1}\right)\dots\chi_{s}T\left(\chi_{s}\right)\dots\chi_{k+r}T\left(\chi_{k+r}\right)\exp\left(\hat{h}\right),\label{eq:PositiveOperator}
\end{equation}
where $\chi_{s}=f_{k_{s}}$ or $f_{k_{s}}^{+}$($s=1,\dots,k+r$)
is a complex linear combination of fermionic creation or annihilation
operators. In this expression, there are $k$ pairs of annihilation
operators and $r$ pairs of creation operators in total, except for
the exponential part. Here
\begin{equation}
\hat{h}=\left(\begin{array}{cc}
c^{+} & d^{+}\end{array}\right)\left(\begin{array}{cc}
M & S\\
-\bar{S} & \bar{M}
\end{array}\right)\left(\begin{array}{c}
c\\
d
\end{array}\right),
\end{equation}
where $M$ and $S$ are abitrary $L\times L$ complex square matrices.
Obviously $T\left(\hat{h}\right)=\hat{h}$, and $\hat{h}$ is an representation
of the $gl\left(L,\mathbb{H}\right)$ Lie algebra $\left(\begin{array}{cc}
M & S\\
-\bar{S} & \bar{M}
\end{array}\right)$. Operators with the form $\chi_{s}T\left(\chi_{s}\right)$ are elements of the representation of the
$so\left(2L,\mathbb{H}\right)$ Lie algebra. They and their commutators span the whole representations of the
$so\left(2L,\mathbb{H}\right)$ Lie algebra, the complexification of which is the $so\left(4L,\mathbb{C}\right)$ Lie algebra.
So $K_{O}$ has the same dimension as $V_{O}$. $K_{O}$ is a proper cone in $V_{O}$.

For any arbitrary $2L\times2L$ square matrix $X$, we have $e^{f^{+}Xf}f_{k}^{+}=f_{\left(e^{X}k\right)}^{+}e^{f^{+}Xf}$.
So when we take the product of any two operators of the form in Eq.~(\ref{eq:PositiveOperator}),
the exponential parts can always be moved to the right end. Since
the $GL\left(L,\mathbb{H}\right)$ Lie group is an exponential Lie
group, the two exponential parts merge into a single exponential.
After moving the exponential part to the right end, the part expressed
by $\chi_{s}T\left(\chi_{s}\right)$ is still time-reversal invariant
and its form remain unchanged. Thus we have proved that the cone $K_{O}$
is closed under operator addtion and operator multipilication, i.e.,
it is a semiring. Actually it is a $Z$-graded semiring, where $Z$
corresponds to the change of the total particle number divided by $2$.

Another interesting property of $K_{O}$ is that all the operators
in $K_{O}$ has non-negative trace. A proof based on Wick's theorem
can be found in Ref.~\onlinecite{wei_semigroup_2024}.

In the linear space of the states with odd particle numbers, $T^{2}=-1$.
In the linear space of the states with even particle numbers, $T^{2}=1$.
Consider the real vector space $V$ spanned by time-reversal symmetric
states $\left|\psi\right\rangle $ with even particle numbers, $T\left|\psi\right\rangle =\left|\psi\right\rangle $.
The subspace of the Fock space with even particle numbers can be seen as a complexification of $V$. $V$
is spanned by all the states of the form $\left(-1\right)^{n}\left|k_{1}T\left(k_{1}\right)\dots k_{s}T\left(k_{s}\right)\dots k_{n}T\left(k_{n}\right)\right\rangle $($s=1,\dots,n$),
where $k_{1}$, $T\left(k_{1}\right)$, ..., $k_{n}$, $T\left(k_{n}\right)$
are a set of linearly independent vectors. Define the convex cone $K$
as the set of non-negative linear combinations of such states. Define
the convex cone $K_{2n}$ as the subset of $K$ with total particle
number $N=2n$. Obviously any two elements in $K_{2n}$ have non-negative
overlap. $K_{2n}$ has the same dimension as the real vector space
$V_{2n}$ spanned by time-reversal symmetric states with total particle
number $N=2n$. $K_{2n}$ is a proper cone in $V_{2n}$.

We say the operators in $K_{O}$ have time-reversal positivity, and
the states in $K$ have Fock space time-reversal positivity.

Just like the case for Majorana reflection positivity for operators\citep{wei_majorana_2016},
the non-negative property of the trace can be used to avoid the sign
problem in auxiliary field quantum Monte Carlo(AFQMC) simulations.
The sampling weights in this case are the traces of time-reversal
positive operators\citep{wei_semigroup_2024}, which can be shown
using the polar decomposition for related semigroups. At zero temperature,
the non-negative property of the overlaps between states with Fock
space time-reversal positivity can be used to avoid the sign problem
in projector quantum Monte Carlo(PQMC) simulations.

\section{Lieb's theorem in time-reversal symmetric case}

The proof of the Lieb's theorem in time-reversal symmetric case exactly follows
the method in Ref.~\onlinecite{wei_liebs_2025}.

Consider a non-hermitian quantum lattice fermion model with time-reversal
symmetry on an arbitrary finite-size lattice $\Lambda$($L=\left|\Lambda\right|$
is the lattice size) defined by
\begin{align}
H & =H_{0}+H_{U},\\
H_{0} & =-f^{+}T_{0}f,\\
H_{U} & =\sum_{i\in\Lambda}U_{i}\left(c_{i}^{+}c_{i}-\frac{1}{2}\right)\left(d_{i}^{+}d_{i}-\frac{1}{2}\right),
\end{align}
$T_{0}$ is the matrix of hopping coefficients, $\left(i\sigma_{y}\otimes I_{L}\right)T_{0}\left(-i\sigma_{y}\otimes I_{L}\right)=\bar{T_{0}}$.
We assume $U_{i}<0$ for all $i\in\Lambda$, which means the interaction
is strictly attractive. We study the uniqueness of the ground state
for this model with particle number $N=2n$.

For any element $\left|\Psi\right\rangle$
in $K_{2n},$ we can define its kernel by
\begin{equation}
\ker\left|\Psi\right\rangle =\left\{ k_{1}\in\mathbb{C}^{2L}\middle|\left\langle k_{1}T\left(k_{1}\right)\dots  k_{n}T\left(k_{n}\right)\middle|\Psi\right\rangle \equiv 0,\forall k_{2},\dots,k_n\in\mathbb{C}^{2L}\right\}.
\end{equation}
Define the range of $\left|\Psi\right\rangle$ by $\operatorname{range}\left|\Psi\right\rangle
=\left(\ker\left|\Psi\right\rangle\right)^{\perp}$.
Given any linear subspace $L_{0}$ of $\mathbb{C}^{2L}$, if $\forall k\in L_{0}$
implies $T\left(k\right)\in L_{0}$, we say $L_{0}$ is time-reversal
symmetric. For any time-reversal symmetric linear subspace $L_{0}$, we
can define a face of $K_{2n}$ by
\begin{equation}
F_{L_{0}}=\left\{ \left|\Psi\right\rangle \in K_{2n}\middle|\operatorname{range}\left|\Psi\right\rangle \subset L_{0}\right\} .
\end{equation}
One can show that all the faces of $K_{2n}$ take this form.

For any positive real number $\beta$, the eigenvector corresponding
to the spectral radius of $\exp\left(-\beta H\right)$ is the ground
state. Let $\beta=M\tau$, where $M$ is a positive integer. Carry
out the following Trotter-Suzuki decomposition:
\begin{equation}
\exp\left(-\beta H\right)=\left[\exp\left(-\tau H_{0}\right)\exp\left(-\tau H_{U}\right)\right]^{M}+O\left(M\tau^{2}\right).
\end{equation}
The error term $O\left(M\tau^{2}\right)$ will disappear as $M$ approaches
infinity. Then for each $\exp\left(-\tau H_{U}\right)$ in this expression,
we can carry out the Hubbard-Stratonovich transformation:
\begin{align}
 & \exp\left[-\tau U_{i}\left(c_{i}^{+}c_{i}-\frac{1}{2}\right)\left(d_{i}^{+}d_{i}-\frac{1}{2}\right)\right]\nonumber \\
= & \exp\left(\frac{\tau U_{i}}{4}\right)\sqrt{\frac{1}{2\pi\tau\left|U_{i}\right|}}\int_{-\infty}^{+\infty}\exp\left[-\frac{x_{i}^{2}}{2\tau\left|U_{i}\right|}-\left(c_{i}^{+}c_{i}+d_{i}^{+}d_{i}-1\right)x_{i}\right]dx_{i}\\
= & \exp\left(-\frac{\tau U_{i}}{4}\right)\sqrt{\frac{1}{2\pi\tau\left|U_{i}\right|}}\int_{-\infty}^{+\infty}\exp\left[-\frac{x_{i}^{2}}{2\tau\left|U_{i}\right|}-i\left(c_{i}^{+}c_{i}-d_{i}^{+}d_{i}\right)x_{i}\right]dx_{i}.
\end{align}
$\exp\left(-\beta H\right)$ can be expressed by multiplications and
non-negative integrations of the operators with the form $\exp\left(\hat{h}\right)$.
Thus $\exp\left(-\beta H\right)$ belongs to the convex cone generated
by the $GL\left(L,\mathbb{H}\right)$ Lie group and is a $K_{2n}$-non-negative
matrix.

To show the ground state is unique in the subspace with $N=2n$ particles,
we only have to show that $\exp\left(-\beta H\right)$
is a $K_{2n}$-irreducible $K_{2n}$-non-negative matrix. To this purpose, we have to show that any
$\exp\left(-\beta H\right)$-invariant face $F\subseteq K_{2n}$ is a trivial face.
This can be done if we can prove that for
any two non-zero extreme vectors $\left(-1\right)^{n}\left|k_{1},T\left(k_{1}\right),\dots,k_{n},T\left(k_{n}\right)\right\rangle$
and $\left(-1\right)^{n}\left|k^{\prime}_{1},T\left(k^{\prime}_{1}\right),\dots,
k^{\prime}_{n},T\left(k^{\prime}_{n}\right)\right\rangle$,
we have $\left\langle k^{\prime}_{1},T\left(k^{\prime}_{1}\right),\dots,
k^{\prime}_{n},T\left(k^{\prime}_{n}\right)\right|e^{-\beta H}\left|k_{1},T\left(k_{1}\right),\dots,k_{n},T\left(k_{n}\right)\right\rangle>0$.

According to the Trotter-Suzuki decomposition and the Hubbard-Stratonovich transformation, the action of $\exp\left(-t H\right)$($t\in\left[0,\beta\right]$) on any non-zero
extreme vector $\left|k_{1},T\left(k_{1}\right),\dots,k_{n},T\left(k_{n}\right)\right\rangle$ in $K_{2n}$
can be seen as an It\^o process on $\mathbb{R}\times\mathbb{C}^{2L\times 2n}$. It
gives a parameterized collection of random variables $\left\{\left(R_{t}, \Psi_{t}\right)\right\} _{t\in\left[0,\beta\right]}$, where $R_{t}$ is defined on
on $\mathbb{R}$, and $\Psi_{t}$ is defined on $\mathbb{C}^{2L\times 2n}$. They satisfy the following stochastic
differential equations:
\begin{align}
&dR_{t}=\sum_{i\in\Lambda}\sqrt{\frac{\left|U_{i}\right|}{2}}dB^{i}_{t},\\    
&d\Psi_{t}=T_{0}\Psi_{t}dt-\sum_{i\in\Lambda}\sqrt{\frac{\left|U_{i}\right|}{2}}T_{1i}\Psi_{t}dB^{i}_{t}-\sum_{i\in\Lambda}\sqrt{\frac{\left|U_{i}\right|}{2}}T_{2i}\Psi_{t}dB^{i+L}_{t},\\
&R_{0}=0,\Psi_{0}=\left(k_{1},T\left(k_{1}\right),\dots,k_{n},T\left(k_{n}\right)\right).
\end{align}
where the matrices $T_{1i}$
satisfies $f^{+}T_{1i}f=c^{+}_{i}c_{i}+d^{+}_{i}d_{i}$, $T_{2i}$
satisfies $f^{+}T_{2i}f=i\left(c^{+}_{i}c_{i}-d^{+}_{i}d_{i}\right)$,
$\mathbf{B}_{t}=\left(B^{1}_{t},B^{2}_{t},\dots,B^{2L}_{t}\right)^{T}$
is the $2L$-dimensional standard Brownian motion. Rewrite
this It\^o form of stochastic differential equations into Stratonovich
form:
\begin{align}
&dR_{t}=\sum_{i\in\Lambda}\sqrt{\frac{\left|U_{i}\right|}{2}}dB^{i}_{t},\\    
&d\Psi_{t}=V_{0}\left(\Psi_{t}\right)dt+\sum_{i\in\Lambda}V_{1i}\left(\Psi_{t}\right)\circ dB^{i}_{t}+\sum_{i\in\Lambda}V_{2i}\left(\Psi_{t}\right)\circ dB^{i+L}_{t},\\
&R_{0}=0,\Psi_{0}=\left(k_{1},T\left(k_{1}\right),\dots,k_{n},T\left(k_{n}\right)\right).
\end{align}
where $V_{0}\left(\Psi_{t}\right)=\left(T_{0}+\frac{1}{4}\sum_{i\in\Lambda}U_{i}T_{1i}^{2}+\frac{1}{4}\sum_{i\in\Lambda}U_{i}T_{2i}^{2}\right)\Psi_{t}$,
$V_{1i}\left(\Psi_{t}\right)=-\sqrt{\frac{\left|U_{i}\right|}{2}}T_{1i}\Psi_{t}$,
$V_{2i}\left(\Psi_{t}\right)=-\sqrt{\frac{\left|U_{i}\right|}{2}}T_{2i}\Psi_{t}$.

According to the corollary of the Stroock--Varadhan support theorem,
the solutions to the ordinary differential equations
\begin{align}
&dr_{t}=\sum_{i\in\Lambda}\sqrt{\frac{\left|U_{i}\right|}{2}}dh^{i}_{t},\\
&d\psi=V_{0}\left(\psi\right)dt+\sum_{i\in\Lambda}V_{1i}\left(\psi\right)dh^{i}++\sum_{i\in\Lambda}V_{2i}\left(\psi\right)dh^{i+L},\\
&r\left(t=0\right)=0,\psi\left(t=0\right)=\left(k_{1},T\left(k_{1}\right),\dots,k_{n},T\left(k_{n}\right)\right)
\end{align}
decide the topological support of $\left(R,\Psi\right)$, where $\mathbf{h}\in W^{1,2}_{0}\left(\left[0,\beta\right],\mathbb{R}^{2L}\right)$.
If the Lie semigroup generated by matrices $T_{0}+\frac{1}{4}\sum_{i\in\Lambda}U_{i}T_{1i}^{2}+\frac{1}{4}\sum_{i\in\Lambda}U_{i}T_{2i}^{2}$,
$\pm\sqrt{\frac{\left|U_{i}\right|}{2}}T_{1i}$ and $\pm\sqrt{\frac{\left|U_{i}\right|}{2}}T_{2i}$ is exactly the $GL\left(L,\mathbb{H}\right)$ Lie group,
$\psi\left(t=\beta\right)$ can reach any point in the orbital $\left\{g\psi\left(t=0\right)\middle|g\in GL\left(L,\mathbb{H}\right)\right\}$ as $\mathbf{h}$ changes. In this case
\begin{align}
&\left\langle k^{\prime}_{1},T\left(k^{\prime}_{1}\right),\dots,
k^{\prime}_{n},T\left(k^{\prime}_{n}\right)\right|e^{-\beta H}\left|k_{1},T\left(k_{1}\right),\dots,k_{n},T\left(k_{n}\right)\right\rangle\nonumber\\
=&\mathbb{E}\left(\left\langle k^{\prime}_{1},T\left(k^{\prime}_{1}\right),\dots,
k^{\prime}_{n},T\left(k^{\prime}_{n}\right)\right|e^{R_{\beta}}\left|\Psi_{\beta}\right\rangle\right)>0
\end{align}
for any two non-zero extreme vectors $\left|k_{1},T\left(k_{1}\right),\dots,k_{n},T\left(k_{n}\right)\right\rangle$
and $\left|k^{\prime}_{1},T\left(k^{\prime}_{1}\right),\dots,
k^{\prime}_{n},T\left(k^{\prime}_{n}\right)\right\rangle$.

To apply our results, a typical choice of $T_{0}$ should fulfill the
following two conditions: (1) $T_{0}$ is connected, i.e., starting from any
site $i_{1}\in\Lambda$ and flavor $\sigma_{1}=c$ or $d$, a particle can hop to all the sites $i_{k}\in\Lambda$ and flavors $\sigma_{k}$
step by step through bonds with $T_{\sigma_{m+1}i_{m+1},\sigma_{m}i_{m}}\neq0$($m=1,\dots,k-1$).
(2) Starting from any site $i_{1}=i\in\Lambda$ and flavor $\sigma_{1}=c$,
a particle can hop to the same site $i_{k}=i$ but the other flavor $\sigma_{k}=d$
step by step thought bonds with $T_{\sigma_{m+1}i_{m+1},\sigma_{m}i_{m}}\neq0$ indirectly, i.e., instead of through the bond $T_{\sigma^{\prime}i,\sigma i}$ directly. With this choice of $T_{0}$, one can generate the whole $gl\left(L,\mathbb{H}\right)$
Lie algebra by taking commutators with $T_{1i}$ and $T_{2i}$.
Hubbard model with Rashba term can serve as a good example of our theorem.

\begin{acknowledgments}
We would like to thank Chong Zhao for helpful discussion.
\end{acknowledgments}

\subsection*{Conflict of Interest}
The authors have no conflicts to disclose.

\section*{DATA AVAILABILITY}
Data sharing is not applicable to this article as no new data were created or analyzed in this study.

\section*{REFERENCES}
\bibliography{TRPositivity}

\end{document}